\documentclass[aps,pra,twocolumn,10pt,superscriptaddress,showpacs,nofootinbib,longbibliography]{revtex4-1}
\usepackage{hyperref,xcolor,graphicx,enumitem,array}
\newcommand{\QUOTE}[2]{\begin{itemize}\item[]\emph{#1} \end{itemize}}
\newcommand{\Quote}[1]{\begin{itemize}\item[]#1\end{itemize}}

\newcommand{\tabDEMO}{
\begin{table}[b]
\caption{\label{tab:demo}{Descriptive information about the universities in the database we created to solicit research participation, and those which are represented in our study. Note that some universities were neither an HBCU, an HSI, nor a PWI.}}
\begin{ruledtabular}
\begin{tabular}{lrrrr}
& & \multicolumn{3}{c}{Population} \\ \cline{3-5}
Classification & Database & Optics & Electronics & Total\\ \hline
Public & 87 & 10 & 10 & 17\\
Private not-for-profit & 67 & 6 & 6 & 10\\ 
Doctoral Universities & 35 & 7  & 7 & 11\\
Master's Universities & 57 & 5 & 3 & 8\\
Baccalaureate Colleges & 62 & 4 & 6 & 8\\ 
Women's Colleges & 8 & 0 & 2 & 2\\
HBCUs & 7 & 0 & 0 & 0\\
HSIs & 44 & 3 & 2 & 4\\
PWIs & 63 & 12 & 10 & 17\\ \hline
Total & 154 & 16 & 16 & 27
\end{tabular}
\end{ruledtabular}
\end{table}
}

\newcommand{\tabCOURSES}{
\begin{table}[b]
\caption{\label{tab:courses}{Characteristics of the courses represented in our study. Small, medium, and large courses had enrollments of fewer than 10, 10 to 35, or over 50 students, respectively. Upper-division labs are those in which students are in their third or higher year of study, and lower-division courses are those in which students are in their first or second year. Mixed-division courses include all levels of students.}}
\begin{ruledtabular}
\begin{tabular}{lcc}
& \multicolumn{2}{c}{Courses} \\ \cline{2-3}
			& Optics	& Electronics \\
Characteristic	& $(N=19)$ & $(N=16)$ \\ \hline
Small		& 42\%	& 13\%\\
Medium		& 47\%	& 69\%\\
Large		& 11\%	& 19\%\\
Lower-division		& 16\%		& 13\%\\
Mixed-division		& 0\%		& 25\%\\
Upper-division		& 79\%		& 63\%\\
Graduate level		& 5\%		& 0\%\\
Minority physics majors	& 0\%	& 13\% \\
Majority physics majors	& 42\%	& 25\%\\
Exclusively physics majors	& 58\%		& 63\%\\
Non-intro lab(s) prerequisite	& 95\%	& 44\% \\
\end{tabular}
\end{ruledtabular}
\end{table}
}

\newcommand{\tabCODES}{
\begin{table*}[!t]
\renewcommand{\arraystretch}{1.5}
\caption{\label{tab:codes}Coding scheme. Code categories and subcategories appear in bolded and italicized font, respectively.}
\begin{ruledtabular}
\begin{tabular}{>{\textbf\bgroup}p{2.25cm}<{\egroup}*{3}{p{4.75cm}}}
\multicolumn{1}{l}{Code} & Definition & Optics examples & Electronics examples  \\ \hline
Make measurements
& Type of equipment used in either the measurement or physical system \emph{apparatus}, the type of \emph{raw data} output by the measurement equipment, or \emph{general} comments about the process of making a measurement.
& ``We use photodiodes to measure the beam that transmits through the plastic puck." ``[Students measure] voltage from the photodiode as a function of the angle at which the [puck] has been rotated." ``We have them measure reflection and transmission off of some surface."
& ``In terms of the measurements, our primary tool that we use for almost everything is the oscilloscope." ``They're measuring  phase shift as a function of frequency." ``First they do a tutorial and then ideally there would be time for them to actually build the circuit and measure the signals." \\
Construct models
& Principles, concepts, parameters, limitations, simplifications, or assumptions related to \emph{models of the photodiode measurement system},  \emph{models of the op-amp circuit}, or \emph{models of any other aspect of the experiment}. Additionally, the \emph{distinction between physical and measurement system models.}
& ``We assume that photodiode has flat response over visible light range." ``They've gotta know, basically, constructive and destructive interference."  ``When you read a voltage, it could mean two things: that your instrument is doing something  or your physical system is doing something."
& ``They also have been taught the limitations of that model: that it only works between the rails." ``We know the bandwidth of the oscilloscope is really large." ``The distinction between measurement apparatus and sort of physical system is one that can get glossed over." \\
Make comparisons
& \emph{Interpreting or analyzing data}, \emph{making or testing predictions}, or \emph{general} comments about the process of comparing data to predictions or expectations. 
& ``They'll average the measurements, they'll determine standard deviations of the measurements." ``They know what to expect, that there's going to be this magic angle where the reflection goes away."  ``They compare the capacitance they measure with the datasheet."
& ``For the Bode plot, they have to learn how to convert voltage into [decibels]." ``The one thing they'll predict is, take your two resistors and calculate a gain." ``I do have them, in some sense, making a comparison, because they check it with the gain number they've calculated." \\
Propose causes
&  Students, instructors, or both propose causes for discrepancies between data and predictions. There are no \emph{a priori} subcategories for this category.
& ``That's how we actually evaluate the report. It's more about, `Can they explain the disagreement?'" ``I make them write down what are the sources of error."
& ``I will try to get them to reason through what it is they might be seeing." ``I try to propose in a general way things that they ought to check out, that maybe they haven't thought about." \\
Enact revisions
& Instances or types of \emph{apparatus revisions}, \emph{model revisions}, or \emph{other iterations} that students enact during the activity.  Additionally, scenarios in which there is  \emph{no opportunity for revision}. 
& ``They definitely do enact revisions to the equipment through the alignment process." ``Some students will try to model the imperfect beam; they'll deviate from pure Gaussian beam propagation." ``They can always go back so they can take more data."  ``[The purpose of the lab] is not to have them enact revisions."
& ``They go back and rebuild the circuit because a lot of times they'll do a miswire." ``Most of them  figure out that the gain equation only works when you reach a certain value." ``There are very few students who get these things to work the first time." ``I doubt that we ever have the time to really iterate." \\
Learning goals
& A learning goal or outcome \emph{related to modeling} or \emph{related to anything else}.
& ``I definitely think constructing models for their equipment is something that I really focus on; I think that's very important."  ``This course is one of the important to develop their writing skills."
& ``I would like them to be able to build an op-amp circuit from a schematic, debug it, make it work."  ``It's important in lab courses for them to come out of there feeling a kind of sense of empowerment." \\
\end{tabular}
\end{ruledtabular}
\end{table*}
}

\newcommand{\tabMODELS}{
\begin{table}[b]
\caption{\label{tab:models}Important components of models of photodiodes and op-amps or op-amp circuits, as identified by optics and electronics lab instructors, respectively.}
\begin{ruledtabular}
\begin{tabular}{lcc}
& \multicolumn{2}{c}{Instructors} \\ \cline{2-3}
& Optics & Electronics  \\
Model component &  $(N=19)$ & $(N=16)$\\ \hline
Output linearly proportional to input 	& 74\%	& 88\% \\
Finite dynamic range				& 68\%	& 63\% \\
Finite bandwidth or response time	& 42\% 	& 63\% \\
Black box 						& 42\%	& 56\% \\
Internal structure of device		& 42\%	& 44\% \\
Background noise or dc offset		& 37\%	& 25\% \\
Diagrams or schematics 			& 12\% 	& 69\% \\
Active area of photodiodes 		& 58\% 	& -- \\
Photodiodes as current source		& 32\%	& -- \\
Finite spectral range of photodiodes 	& 26\% 	& -- \\
Golden rules for op-amp circuits 	& --		& 81\% \\
Feedback in op-amp circuits 		& -- 		& 50\% \\
Finite input impedance of op-amps	& --		& 31\%
\end{tabular}
\end{ruledtabular}
\end{table}
}

\newcommand{\figFRAMEWORK}{
\begin{figure*}[t]
\includegraphics[width=0.667\textwidth]{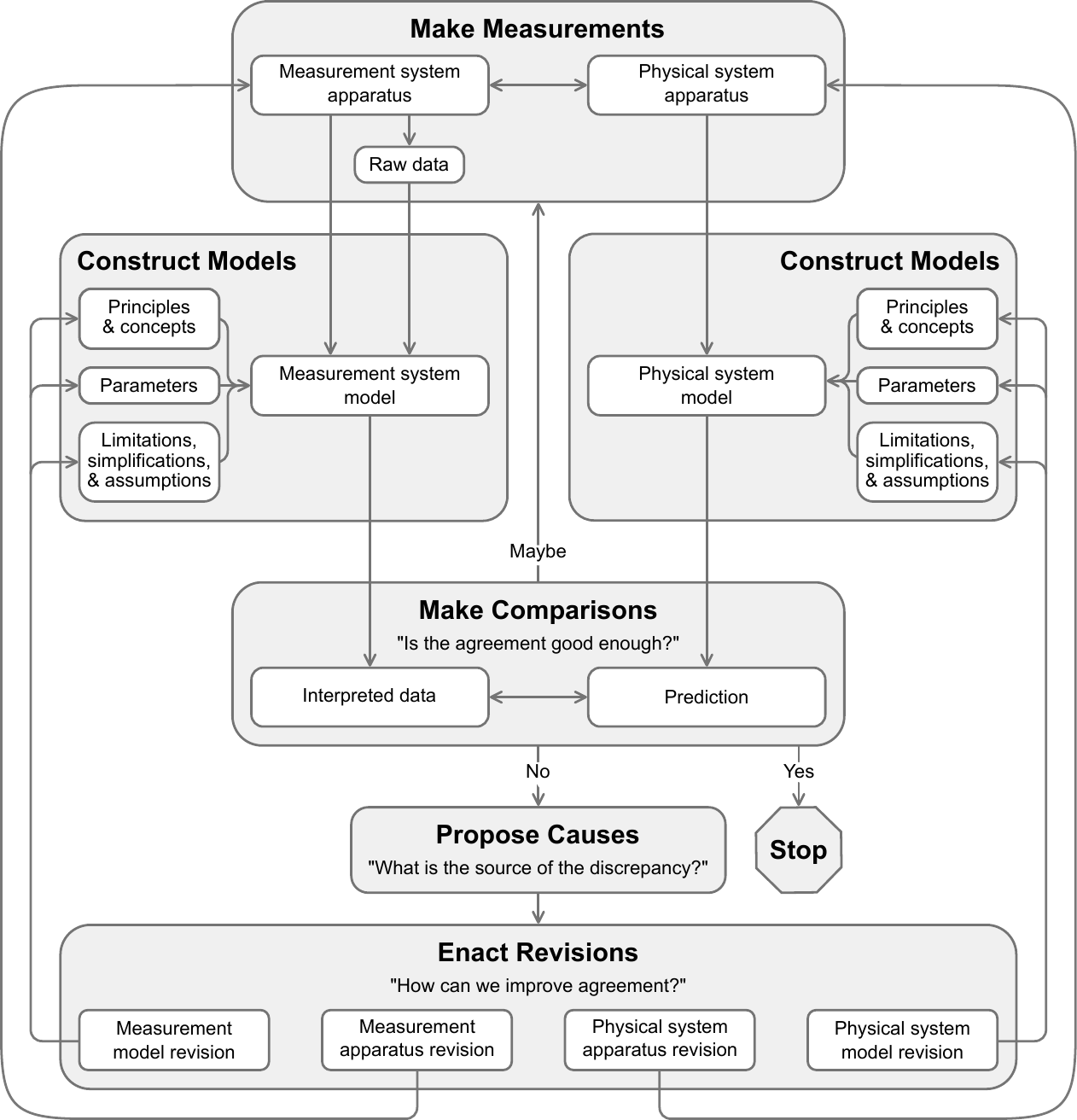}
\caption{\label{fig:framework}{Modeling Framework for Experimental Physics}. This framework describes the recursive and nonlinear process of modeling. The six gray boxes each correspond to a distinct subtask: making measurements, constructing models of the measurement system, constructing models of the physical system, comparing data to predictions, proposing causes for any disagreements between data and predictions, and {enacting} revisions to models or apparatus of either the measurement or physical system. The diagram here differs slightly from that of Ref.~\cite{Zwickl2015a}. {Most notably,} this version includes a ``Maybe" pathway from making comparisons to making measurements. {Bold phrases indicate aspects of the framework that informed our \emph{a priori} coding scheme.}}
\end{figure*}
}

\newcommand{\figCHARTS}{
\begin{figure*}
\includegraphics[width=\textwidth]{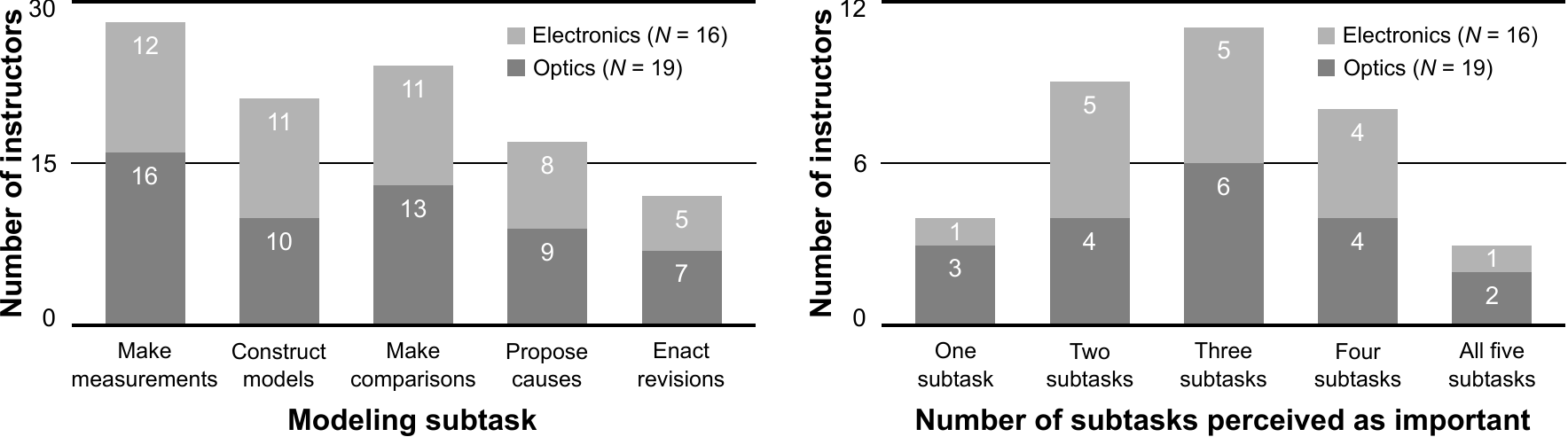}
\caption{\label{fig:charts} (Left) Stacked bar chart showing the number of optics instructors and electronics instructors who identified a particular modeling subtask as an important learning outcome for students. (Right) Stacked bar chart showing the number of instructors who identified a particular number of modeling subtasks as important for students to learn. For both charts, the dark grey bars on the bottom of each stack correspond to optics instructors and the light gray ones on the top correspond to electronics instructors.}
\end{figure*}
}

\begin{document}


\title{{Characterizing lab instructors' self-reported learning goals
\\to inform development of an experimental modeling skills assessment}}

\author{Dimitri R. Dounas-Frazer}
\email{dimitri.dounasfrazer@colorado.edu}
\affiliation{Department of Physics, University of Colorado Boulder, Boulder, CO 80309, USA}
\affiliation{JILA, National Institute of Standards and Technology and University of Colorado Boulder, Boulder, CO 80309, USA}

\author{Laura R\'ios}
\affiliation{Department of Physics, University of Colorado Boulder, Boulder, CO 80309, USA}
\affiliation{JILA, National Institute of Standards and Technology and University of Colorado Boulder, Boulder, CO 80309, USA}

\author{Benjamin Pollard}
\affiliation{Department of Physics, University of Colorado Boulder, Boulder, CO 80309, USA}
\affiliation{JILA, National Institute of Standards and Technology and University of Colorado Boulder, Boulder, CO 80309, USA}

\author{Jacob T. Stanley}
\altaffiliation[Current address: ]{BioFrontiers Institute, University of Colorado, Boulder, CO 80309, USA}
\affiliation{Department of Physics, University of Colorado Boulder, Boulder, CO 80309, USA}

\author{H. J. Lewandowski}
\affiliation{Department of Physics, University of Colorado Boulder, Boulder, CO 80309, USA}
\affiliation{JILA, National Institute of Standards and Technology and University of Colorado Boulder, Boulder, CO 80309, USA}

\date{\today}

\begin{abstract}
The ability to develop, use, and refine models of experimental systems is a nationally recognized learning outcome for undergraduate physics lab courses. However, no assessments of students' model-based reasoning exist for upper-division labs. This study is the first step toward development of modeling assessments for optics and electronics labs. {We interviewed 35 lab instructors about the ways they incorporate modeling in their courses, and we used their self-reported learning goals and activity designs to identify test contexts and objectives that are likely relevant across many institutional settings.} The study design was informed by the Modeling Framework for Experimental Physics, {which} conceptualizes modeling as consisting of multiple subtasks: making measurements, constructing system models, comparing data to predictions, proposing causes for discrepancies, and enacting revisions to models or apparatus. We found that each modeling subtask was identified by {multiple instructors} as an important learning outcome for their course. Based on these results, we argue that test objectives should include probing students' competence with {most} modeling subtasks, and test items should be designed to elicit students' justifications for choosing particular modeling pathways. In addition to discussing these and other implications for assessment, we also identify future areas of research related to the role of modeling in optics and electronics labs.
\end{abstract}

\maketitle

\section{\label{sec:introduction}Introduction}
The ability to develop, use, and refine models of the natural world is nationally recognized as an important learning outcome for physics students at all levels~\cite{JTUPP2016,NRC2012,NGSS2013}, including those in undergraduate physics labs~\cite{AAPT2015}. Koponen~\cite{Koponen2007} argues that modeling is an inherently empirical process and therefore, ``in order to learn to use models in physics, it is crucial to recognize that this learning needs to be done in the context of experiments and experimentation." (p.~767). The role of models and modeling in  physics education has been a major focus of research for about three decades. {In the late 1980s,} Hestenes and Halloun~\cite{Hestenes1987,Halloun1987} laid the groundwork for a model-centered curriculum that {is now} known as Modeling Instruction~\cite{Brewe2008,Megowan-Romanowicz2011}, a widely used pedagogy for introductory physics at the high school and university levels. Since then, several other introductory courses and curricula have been developed to engage students in the iterative process of creating and revising models~\cite{Etkina2007,Buffler2008,Steinberg2008a,Steinberg2008b,Gandhi2016,Vonk2017}. At the upper-division level, the Advanced Lab~\cite{Zwickl2012,Zwickl2013,Zwickl2014} and Electronics Lab~\cite{Lewandowski2015,Stanley2017} at the University of Colorado Boulder have both been transformed to emphasize model-based reasoning. Despite this sustained and multifaceted interest in modeling, few standardized assessments of students' physics modeling abilities exist.

{The} National Research Council (NRC) has called for additional research in lab courses~\cite{DBER2012}. More specifically, the NRC has called for the development of nationally normed assessments of experimental physics practices {and} increased attention to the development process itself~\cite{NRC2013}. In particular, there is a need to create instruments that measure students' competence with modeling. For example, Laverty and Caballero~\cite{Laverty2017arXiv} showed that none of the four most widely used physics concept surveys target the concept of systems and system models or the practice of developing and using models. While there are a few ongoing efforts to assess some aspects of students' model-based reasoning in introductory physics lab and lecture courses~\cite{Kuo2017,McPadden2017,Holmes2015,*Holmes2016}, no modeling assessments exist for upper-division labs. To address these national priorities and needs, we are using a four-phase process to develop standardized and scalable assessments of students' experimental modeling abilities. Here, we report on the first phase: aligning assessment design with lab instructors' self-reported learning goals.

{The four phases of our development plan are (i) establish test objectives, (ii) characterize student navigation of lab practicum-style activities, (iii) create a free-response assessment, and (iv) create a validated multiple-response-style assessment. Students' engagement with the practicum-style activities will inform the types of questions asked on the free-response assessment, and students' written responses to free-response test items will inform multiple-response options on the final version of the assessment. The test objectives will inform design and analysis of all instrument formats. We are using this plan to create two distinct modeling assessments for upper-division electronics and optics lab courses. For each instrument, the Modeling Framework for Experimental Physics~\cite{Zwickl2014perc,Zwickl2015a} (Fig.~\ref{fig:framework}) forms the underlying theoretical basis for our assessment designs.}

Ultimately, we aim to develop instruments that leverage advantages of two approaches to assessment that {have been described recently} in the physics education literature: a Coupled Multiple Response (CMR) format~\cite{Wilcox2014} and an administration model that centralizes data collection and analysis~\cite{Wilcox2016b}. We will use these approaches to create assessments that facilitate nationwide studies of students' {experimental modeling skills} in lab courses. In order to develop instruments tailored to the goals of national deployment and centralized administration, we want to ensure that the objectives and contexts of the assessments will be relevant to the goals and activities of as many lab courses as possible throughout the United States.

Establishing test objectives that are consistent with course learning goals is one of the first steps toward the development of research-based assessments~\cite{Wilcox2015,Engelhardt2009}. However, because modeling is only one of several nationally recommended learning outcomes and consists of multiple subtasks and recursive pathways, it is unclear whether and how these subtasks are prioritized among the major learning goals of a particular course. {Moreover, both the goals and content of upper-division labs can vary widely from one department to the next. Hence, it is important to identify which goals and activities are common for particular types of lab courses.} Survey and interview studies can help clarify instructors' goals~\cite{Coppens2016a,Dounas-Frazer2016b,Dounas-Frazer2017}. Such empirical evidence about course goals {can motivate and inform} corresponding test objectives for research-based assessments. This {reasoning} is precisely the rationale that inspired the present study.

We conducted an exploratory study of instructors' perceptions about the role of models and modeling in two types of courses: optics labs and electronics labs. In this study, we aim to achieve three goals related to the design of modeling assessments, each corresponding to a research question. Our first goal is to identify which modeling subtasks align with instructors' self-reported learning goals. {Doing so} will facilitate creation of test objectives that will be relevant at a national scale. Accordingly, we ask,
\begin{itemize}
\item[Q1.] According to instructors, which subtasks of the Modeling Framework are important for students to learn during (a) optics lab courses and (b) electronics lab courses?
\end{itemize}
Our second goal is to identify common models of photodiodes and operational amplifiers (op-amps) that can be used to contextualize test items in ways that will be familiar to many lab students across the country. We chose to focus on photodiodes and op-amps because, based on our experience teaching and studying lab courses and our participation in professional communities dedicated to lab instruction,\footnote{Namely, the American Association of Physics Teachers (AAPT) Committee on Laboratories and the Advanced Laboratory Physics Association (ALPhA).} they are used in almost all optics and electronics labs, respectively.  Therefore, photodiodes and op-amps will likely feature in future phases of assessment development. In order to better understand {which} types of models students use when working with these common pieces of equipment, we ask,
\begin{itemize}
\item[Q2.] According to instructors, how do students model (a) photodiodes in optics labs and (b) op-amps and op-amp circuits in electronics labs?
\end{itemize}
Our third goal is to determine whether and how modeling assessments should be uniquely tailored to specific lab courses. To this end, we ask,
\begin{itemize}
\item[Q3.] What similarities and differences exist, if any, in the ways that models and modeling manifest in optics labs as compared to electronics labs?
\end{itemize}
To answer these questions, we interviewed 35 lab instructors: 19 optics instructors and 16 electronics instructors. The Modeling Framework informed the development of both the interview protocol and the analysis scheme. We have previously reported results from a preliminary analysis of a subset of these data~\cite{Dounas-Frazer2017arXiv}. Here, we present a more comprehensive analysis. 

\section{\label{sec:background}Background}
We focus on three overlapping areas of study: (A) conceptions of models and modeling in science education, (B) investigations of instructors' understanding and use of modeling, and (C) assessments of students' lab skills and modeling knowledge or ability. We draw from the science education literature broadly, with an emphasis on work in undergraduate physics contexts.

\subsection{Models and modeling in science education}
What is a model, how are models used, and how are they developed? These questions form the basis of a large number of investigations in science education. In a recent overview of the roles of models and modeling in teaching science, Gilbert and Justi~\cite{Gilbert2016a} argue in favor of Knuuttila's~\cite{Knuuttila2011} philosophical interpretation of models as epistemic artifacts, i.e., objects that facilitate knowledge generation. {In the artifactual view, models serve many purposes:} abstracting, idealizing, or representing systems; supporting arguments, explanations, or predictions; or designing experiments or other models. Models are expressed externally through a variety of representational modes, such as equations, simulations, diagrams, three-dimensional objects, or verbal descriptions. Gilbert and Justi distinguish Knuuttila's artifactual view of models from the idea of mental models, popularized in part by Johnson-Laird~\cite{Johnson-Laird1980}. In contrast to externally expressed artifactual models, mental models are internal representations constructed in the mind.  Despite this distinction, Gilbert and Justi argue that mental models play an important role in the process by which artifactual models are created. For example, they propose that internal representations enable the creation of external ones.

Giere~\cite{Giere2009} argues that it is less important to define exactly \emph{what a model is} than to describe \emph{what a model does}. According to Giere, agents intend to use models to represent a part of the world for some purpose. Building on Giere's interpretation of the function of models, Gouvea and Passmore~\cite{Gouvea2017} describe models as knowledge-generating tools used by an agent for a particular epistemic goal. The agent decides what to model and why, and they evaluate and refine their models according to their goal. In this context, valuable models are not necessarily those whose external representations are highly realistic {or map} isomorphically onto reality. Rather, they are those that succeed in their intended uses, such as making accurate predictions~\cite{Gilbert2016a,Gouvea2017,Koponen2007}. The process through which an agent creates, uses, evaluates, or revises a model is called \emph{modeling}.

In physics education, the purpose of modeling is often to solve problems, improve conceptual understanding, sense-make, or generate accurate descriptions, explanations, or predictions about physical systems~\cite{Etkina2007,Brewe2008,Russ2017,Megowan-Romanowicz2011}. {Koponen}~\cite{Koponen2007} argues that physicists commonly use models as tools for conceiving and creating systems that can be explored experimentally.  According to Koponen, the empirical reliability of a model is evaluated through a process of matchmaking. Here ``matchmaking" entails establishing a threefold match between simple models and the more sophisticated theories in which they are nested, model predictions and experimental data, and the models and the phenomena themselves. Koponen describes the process of matching models and phenomena as bi-directional; it includes both fitting models to phenomena and altering phenomena {to} fit models by, e.g., designing an apparatus {to} isolate and observe a particular effect.

Finally, the modeling process is facilitated by various kinds of knowledge, including domain knowledge. Domain knowledge consists of the principles and concepts relevant to the target system being studied. In the case of photodiodes and op-amps, domain knowledge may include principles like conservation of charge and concepts like charge carriers; it may also include a general understanding of electric circuits, including the behavior {of} transistors, diodes, or current sources.

In a study of middle school students' modeling practices, Ruppert et al.~\cite{Ruppert2017} found that domain knowledge played a significant role in students' ability to construct biological system models that accounted for all available evidence. Other work found that high school science students' ability to construct models was dependent on their domain knowledge~\cite{Fortus2016}, and that students evaluate their models based on the models' ability to explain evidence presented during class as well as the students' prior knowledge from outside of class~\cite{Stewart2005}. {Furthermore, in the context of an experimental optics task, Zwickl et al.~\cite{Zwickl2015a} argued that, ``When a lab activity utilizes concepts that are largely outside of studentsÕ prior knowledge, it has a significant impact on how they engage in the laboratory." (p.~11).} These findings suggest that constructing models, making comparisons, {making predictions}, and likely other modeling subtasks are dependent upon one's domain knowledge.

\subsection{Instructors' understanding and use of modeling}
\label{sec:instructors}

Previous research on science instructors' understanding and use of models and modeling has focused primarily on preservice or practicing K-12 teachers. {Many such studies aim to characterize teachers' metamodeling knowledge, i.e., their understanding of the nature and purposes of models~\cite{Krell2016,Fazio2012,Justi2003,VanDriel1999,Crawford2004,Davis2010}. For example,} Krell and Kr\"uger~\cite{Krell2016} found that biology teachers thought of models as idealized depictions used to show or explain something, {but} not as research tools. Davis et al.~\cite{Davis2010} found that preservice K-8 teachers were more likely to understand the explanatory, rather than the predictive, purpose of models. Along similar lines, in a study by Van Driel and Verloop~\cite{VanDriel1999}, biology, chemistry, and physics teachers typically defined models as simplified reproductions of reality, {rarely indicating} that models could be used to make predictions.

Other research has focused on interventions geared toward improving K-12 teachers' understanding of models or supporting their use of models in the classroom~\cite{Megowan-Romanowicz2011,Crawford2004,Davis2010,Soulios2016,Papaevripidou2012,Schwarz2009,Windschitl2008b}. For example, Windschitl et al.~\cite{Windschitl2008b} developed a system of learning activities for preservice teachers that helped them incorporate testing and revising explanatory models into their goals for science learning. And, according to Megowan-Romanowicz~\cite{Megowan-Romanowicz2011}, over 3,000 teachers have participated in training for Modeling Instruction. Taken together, this literature~\cite{Krell2016,Fazio2012,Justi2003,VanDriel1999,Crawford2004,Davis2010,Megowan-Romanowicz2011,Crawford2004,Davis2010,Soulios2016,Papaevripidou2012,Schwarz2009,Windschitl2008b} paints a picture of both need and promise for model-centered teacher professional development.

When it comes to physics instructors at the undergraduate level, the literature is more sparse. Given our focus on optics and electronics lab instructors, two investigations of undergraduate physics instructional practices are relevant to the present work even though they do not focus on models or modeling. First, Coppens et al.~\cite{Coppens2016a} surveyed instructors from Belgian universities about their perceptions of various learning goals for electronics labs. Instructors were given a  list of goals and asked to rank them according to importance. Goals related to learning how to collect, interpret, and analyze data were among those ranked as ``important" or ``very important" by all instructors~\cite{Coppens2016a}. Second, we interviewed electronics instructors from across the United States about their goals and practices related to teaching students how to troubleshoot electric circuits~\cite{Dounas-Frazer2017,Dounas-Frazer2016b}. We found that an underlying belief that ``nothing works the first time" informed instructors' view that troubleshooting is a crucial skill for electronics and experimental physics more broadly~\cite{Dounas-Frazer2016b}. Nevertheless, only half of instructors reported assessing students' troubleshooting ability~\cite{Dounas-Frazer2017}. {While data analysis, data interpretation, and troubleshooting are related to modeling---e.g., we have previously shown that students use model-based reasoning throughout the process of troubleshooting an electric circuit~\cite{Dounas-Frazer2015,Dounas-Frazer2016a}---modeling is different from these skills. We are unaware of any previous studies that specifically probe instructors' use or knowledge of models in undergraduate physics labs}.

\subsection{Relevant existing assessments}

Although science lab courses are not as well studied as lecture or studio courses~\cite{DBER2012}, there are nevertheless several research-based assessments of student learning that are relevant for labs. For example, at the high school level, assessments have been developed to probe students'  knowledge and attitudes about science~\cite{Freedman2002}, perceptions of the lab environment~\cite{Fraser1993}, and their competence with experimental design, data analysis, and other lab skills~\cite{Doran1993}. {These and other high school lab} assessments are outlined in Hofstein and Lunetta's review of labs in science education~\cite{Hofstein2004}. At the undergraduate level, relevant assessments include those that focus on argumentation, experimental design, iteration, and ownership in biology labs~\cite{Deane2014,Gormally2012,Sirum2011,Corwin2015,Hanauer2014} and epistemological, cognitive, and affective aspects of chemistry labs~\cite{Saunders1999,Galloway2015}. For physics labs, course assessments have been developed to probe learning outcomes related to data analysis~\cite{Buffler2001,Day2011} and attitudes about experimental physics~\cite{Zwickl2014e,Wilcox2016}.

Many of these undergraduate lab assessments~\cite{Deane2014,Gormally2012,Sirum2011,Corwin2015,Galloway2015,Buffler2001,Day2011} likely target aspects of learning related to modeling. For example, the Laboratory Course Assessment Survey (LCAS)~\cite{Corwin2015} probes biology students' engagement in various forms of iteration, such as collecting additional data to address questions that arise during investigations or revising data analysis based on feedback. The Meaningful Learning in the Laboratory Instrument (MLLI)~\cite{Galloway2015} probes chemistry students' cognitive and affective experiences in the lab, such as whether they made decisions about what data to collect, considered whether the data made sense, or felt worried or confused about the quality or interpretation of their data. {Last,} the Physics Measurement Questionnaire (PMQ)~\cite{Buffler2001} focuses on physics students' understanding of measurement, in particular the idea that all measurements have inherent uncertainty. However, the iterative collection, analysis, and interpretation of data is only part of the process of developing and revising models. None of these assessments were specifically designed to comprehensively assess students' {experimental modeling skills.}

Outside the context of labs, several previous or ongoing efforts specifically assess students' modeling abilities. In K-12 settings, assessments related to models and modeling tend to focus on students' metamodeling knowledge. In such assessments, understanding of models and modeling is typically broken down into a few categories related to beliefs about the nature, purpose, evaluation, or revision of models. {Examples include the Students' Understanding of Models in Science (SUMS) instrument~\cite{Treagust2002}, the Modeling Test~\cite{Schwarz2005}, and an unnamed assessment developed by Krell et al.~\cite{Krell2014}}. These assessments~\cite{Treagust2002,Schwarz2005,Krell2014} provide useful information about students' metamodeling knowledge, but they do not {directly measure} students' modeling ability.

In undergraduate physics contexts, we are aware of a few assessments that test some aspects of students' modeling ability~\cite{Kuo2017,McPadden2017,Holmes2015,*Holmes2016}. Kuo et al.~\cite{Kuo2017} designed a pen-and-paper instrument to test students' representational competence in an introductory optics lecture course. Similarly, McPadden and Brewe~\cite{McPadden2017} developed the Problem Solving and Representation Use Survey (PSRUS), an online survey that assesses students' choice of representations when reasoning about introductory mechanics, electricity, and magnetism problems. On both instruments, each item consists of a physics problem and multiple representations, including equations, diagrams, and graphs. Students are then prompted to either solve the problem using all available representations (Kuo et al.) or select which representations they would use to solve the problem without actually solving it (McPadden and Brewe). {For labs,} Holmes and Wieman~\cite{Holmes2015,*Holmes2016} are currently developing the Physics Lab Inventory of Critical Thinking (PLIC), a choose-many closed-response survey that aims to measure students' critical thinking while testing a model prediction. In the context of the PLIC, critical thinking is defined as the ability to critique data, determine whether conclusions are supported by evidence, and distinguish signal from noise~\cite{Holmes2018arXiv}. The PLIC is contextualized by a mass-on-spring system and is intended for use in introductory labs. 

The instruments developed by Kuo et al.\ and McPadden and Brewe provide valuable information about the impacts of particular courses on students' ability to use multiple representations when solving problems. And, in the future, the final version of the PLIC may shed light on how students compare model predictions with experimental data. However, external expressions of models are not themselves models~\cite{Gilbert2016a,Gouvea2017}, and matching data to predictions is only one of many aspects of modeling~\cite{Koponen2007}. Therefore, each of these instruments~\cite{Kuo2017,McPadden2017,Holmes2015,*Holmes2016} provides only partial {measures of} students' modeling ability. Moreover, all three assessments are designed to be used in introductory physics courses. At the upper-division level, there are no assessments of students' modeling ability.

In this paper, we report on the first phase of development of modeling assessments in optics and electronics lab courses. The Modeling Framework for Experimental Physics is the theoretical foundation underlying our {assessment design} efforts.

\section{\label{sec:framework}Modeling Framework}
\figFRAMEWORK

Originally developed by Zwickl et al.~\cite{Zwickl2014perc,Zwickl2015a}, the {Modeling Framework for Experimental Physics (Fig.~\ref{fig:framework})} describes the subtasks and cyclic process that physicists employ when refining models and apparatus. The framework conceptualizes the purpose of modeling as achieving ``good enough" agreement between data and predictions or explanations. {Elsewhere, we have provided a detailed review of the Modeling Framework~\cite{Dounas-Frazer2018}. Here, we describe its subtasks, applications, and limitations.}

\subsection{Modeling subtasks}

A diagram of the Modeling Framework is provided in Fig.~\ref{fig:framework}. The diagram can be thought of as a flowchart connecting multiple subtasks: make measurements, construct models of both the measurement and physical systems, make comparisons of data to predictions, propose causes for discrepancies, and enact revisions to models or apparatus in order to improve agreement between data and predictions. The Modeling Framework describes the recursive processes through which systems and system models are brought into alignment with one another by resolving discrepancies between data and predictions. Although it can be read from top to bottom, there is no specific entry point in the diagram. Depending on the context, any subtask could be thought of as the ``start" of a modeling cycle. However, it can be useful to think of making measurements or constructing models as entry points since these subtasks generate data that need to be interpreted or predictions that need to be tested. In this section, we briefly summarize each subtask, starting with the topmost subtask of the diagram: making measurements. Throughout, we use examples from electronics to illustrate abstract concepts.

\textbf{Making measurements} is represented as an interaction between the measurement equipment and the physical system apparatus. In the case of electronics, the measurement equipment includes voltmeters, ohmmeters, probes, {and cables}. The physical system apparatus consists of the breadboard, wires, resistors, capacitors, and other electronic components that compose an electric circuit. The interaction between the measurement and physical system apparatus yields raw data, which is produced by the measurement equipment.

\textbf{Constructing models} appears twice in the framework: once each for the measurement and physical systems. In both cases, the system models incorporate underlying physics principles or concepts (e.g., conservation of charge), relevant physical parameters (e.g., prescribed component values), and simplifying assumptions or limitations (e.g., wires and cables have negligible resistance). In electronics, common external expressions of models include equations, graphs, drawings, diagrams, schematics, data sheets, and computer simulations. The measurement system model is used to interpret the raw data produced by the measurement equipment. For example, when working with ac signals, it is important to know whether a voltmeter is measuring peak-to-peak or root-mean-square voltages. On the other hand, the physical system model is used to make predictions about the behavior of the circuit {itself}.

The scope and complexity of a model is often matched to its intended use. {A} model that treats photodiodes as mechanism-free ``black boxes" that linearly convert light intensity to electric current cannot explain the wavelength-dependent responsivity of a photodiode, but it is practically sufficient for many experimental systems in which photodiodes are illuminated by a monochromatic light source. Indeed, parsimony is a desirable feature of models---that is, a model should be only as complex as necessary for its particular purpose.

\textbf{Making comparisons} refers to the process of determining whether the interpreted data are adequately explained by the model predictions. The level of agreement between data and prediction that counts as ``good enough" is highly dependent upon the context and goals of the experiment (cf. Ref.~\cite{Giere2009}). For example, in {high precision experiments such as searches for the electron electric dipole moment~\cite{ACME2014}}, it is common for researchers to be concerned with very small discrepancies between data and predictions. In contrast, in an undergraduate electronics lab, it is common for instructors to encourage students to move on from one activity to the next after confirming that observed signals are within about 10\% of the predicted values.

Although data analysis is not explicitly represented in the framework, it is nevertheless an implicit part of making comparisons. For instance, when using the Modeling Framework to develop and evaluate model-centered lab activities, Vonk et al.~\cite{Vonk2017} noted that the task of evaluating whether a system's behavior is consistent with a given model involves estimating, calculating, and propagating uncertainties. The type and extent of data analysis required for a given experiment is highly dependent upon how ``good enough" agreement is operationalized in context.

Depending on the outcome of a comparison, the Modeling Framework describes three potential pathways one may pursue after a comparison has been made. If the agreement is good enough for the researcher's needs, then the experiment may come to an end. At this point, a research team may begin to write up their results for publication or a student may start writing their lab report (the ``Yes" pathway). Alternatively, the data may be too noisy in order for a meaningful comparison to be made. In this case, the researcher may repeat the experiment and collect more data in order to improve the statistical precision of the data  (the ``Maybe" pathway). The last possibility is that there is a systematic difference between the data and prediction that must be resolved by revising the models or apparatus in the experiment (the ``No" pathway). {In this sense, the ``Maybe" and ``No" pathways in Fig.~\ref{fig:framework}  roughly correspond to efforts to reduce statistical and systematic uncertainties, respectively.}

Finally, \textbf{proposing causes} and \textbf{enacting revisions} encapsulate the process of suggesting causal explanations for discrepancies and implementing corresponding changes to the experiment in order to bring {data} and models into better alignment. The ultimate goal is to improve agreement between data and predictions. Four types of revisions are included in the Modeling Framework: revising either the apparatus or model of either the measurement or physical system {(cf. Ref.~\cite{Koponen2007})}. After a revision has been made, the modeling process starts over again. {Measurements performed with a new apparatus produce new data, models informed by new assumptions yield new predictions, and these data and predictions are compared until good enough agreement is reached}. {Thus, as Russ and Odden noted, model-based and evidence-based reasoning are intertwined in the framework; evidence is used to construct models, and models are used to inform the search for evidence~\cite{Russ2017}.}

\subsection{{Previous applications of the framework}\label{sec:applications}}
{The Modeling Framework has previously been used to create model-centered activities in an introductory course~\cite{Vonk2017} and in upper-division labs~\cite{Zwickl2012,Zwickl2013,Zwickl2014,Lewandowski2015}. It has also been used to characterize students' modeling approaches in experimental optics~\cite{Zwickl2014perc,Zwickl2015a} and electronics~\cite{Dounas-Frazer2015,Dounas-Frazer2016a} contexts, and to examine students' engagement in modeling during scaffolded model-oriented lab activities in an electronics lab course~\cite{Stanley2017}. Prior research has demonstrated that students engage in a wide variety of modeling subtasks during experimental activities~\cite{Zwickl2015a,Dounas-Frazer2016a}. Further, a few challenging aspects of modeling have been identified. For example, students may not be able to compare data to predictions if there are gaps in their content knowledge~\cite{Zwickl2015a} or if they are unfamiliar with relevant criteria for evaluating the quality of agreement~\cite{Stanley2017}. Similarly, students may not propose or enact revisions to models or apparatus if they are unable to articulate the assumptions of their models~\cite{Zwickl2015a} or if they are not explicitly asked to do so~\cite{Stanley2017}.}

\subsection{Limits of the framework}\label{sec:scope}
The Modeling Framework was developed to describe the process of reaching agreement between experimental data and explanatory or predictive models. Modeling is different from the ``scientific method"~\cite{Windschitl2008a}, and the framework is not meant to describe the initial stages of experimentation such as formulating a research question or designing an investigation. Neither does the framework describe the process of communicating scientific arguments to others. 

Nevertheless, experimental design and scientific argumentation rely on models and modeling. For example, Koponen~\cite{Koponen2007} has argued that, in addition to explaining and predicting phenomena, physicists use models to imagine and construct experimental apparatus that can be used to compare theory to reality. Passmore and Svoboda~\cite{Passmore2012} argue that model-based argumentation arises during the design process, when students must decide what to investigate and how to go about the investigation. They further describe how argumentation may arise ``when students are attempting to use a model to explain a phenomenon" or ``when students are confronted with judging between models or ideas."~(p.~1541). Others have suggested that making arguments is an important feature of model-based instruction~\cite{Windschitl2008a}, and that clear communication of ideas is a hallmark of good {modeling practice}~\cite{Pluta2011}.

Although the Modeling Framework was not intended to describe design or argumentation, some elements of these processes are implicitly embedded in the framework. For example, designing an experimental setup may involve model-based apparatus revisions like those described by the framework. Students likely engage in argumentation when deciding whether to revise a model, an act that involves judging between models. However, the Modeling Framework needs to be combined with other frameworks in order to fully capture these processes. In fact, we have previously used a multiple-framework approach to investigate students' use of model-based reasoning while engaging in a design-related task, namely, troubleshooting~\cite{Dounas-Frazer2016b,Dounas-Frazer2017}.

\tabDEMO
\section{\label{sec:methods}Methods}
This study is an exploratory investigation of instructors' perceptions of the role of modeling in upper-division lab courses. We conducted semistructured interviews with instructors of optics and electronics labs. {Our study was designed to help us identify test objectives for future assessments of students' modeling skills in experimental physics contexts. Our results allow us to determine whether instructors incorporate modeling in their course learning goals and how they aim to engage students in modeling during lab activities.}

{Each interview focused partly on a lab activity of the interviewee's choice. For this part of the interview, optics instructors were asked to choose an activity that incorporated one or more photodiodes. Similarly, electronics instructors were asked to choose an activity that involved an inverting amplifier op-amp circuit. Our rationale for focusing on these pieces of equipment was connected to the scalable nature of the modeling assessments we ultimately aim to create. Modeling always occurs in a particular context, and scalable assessments must be contextualized by systems that are common within the domain of interest. Otherwise, lack of familiarity with apparatus may prevent students from demonstrating their model-based reasoning. Our collective experience teaching and studying lab courses strongly suggests that photodiodes and op-amps are common in optics and electronics labs, respectively. They are therefore good candidates for contextualizing future modeling assessments. Hence, we aim to understand the ways that models and modeling relate specifically to activities that incorporate photodiodes or op-amps. Importantly, this focus does not preclude us from also making statements about the role of modeling in optics and electronics labs more generally.}

{In this section, we describe four parts of our research methods: (A) participant recruitment and demographics, (B) course contexts, (C) data collection, and (D) data analysis. One of our methodological goals was for the results of the investigation to be transferable to a wide range of lab courses. Eisenhart~\cite{Eisenhart2009} recommends that researchers provide ``sufficient detail about the researched context for a person with intimate knowledge of a second context to judge the likelihood of transferability." (p.~56). Accordingly, we provide detailed contextual information for the institutions, departments, courses, and people represented in our study. When describing our coding process, we follow several recommendations made by Hammer and Berland~\cite{Hammer2014}. Namely, we provide definitions and examples of each \emph{a priori} code category, a measure of agreement between two coders, and an example of how the coders resolved a discrepancy in their code assignments.}

\subsection{Participant recruitment and demographics}
To recruit participants, D.R.D.F. and J.T.S. created a database of undergraduate physics programs. The database consisted of three types of programs: the top 15 largest producers of physics bachelor's degree recipients among each of terminal bachelor's, master's, and doctoral programs (45 programs total); all programs at Women's Colleges, Historically Black Colleges and Universities (HBCUs), and Hispanic Serving Institutions (HSIs); and programs chosen randomly from the American Institute of Physics (AIP) roster of physics departments~\cite{AIP2016}. To identify physics programs at Women's Colleges, HBCUs, and HSIs, we cross referenced the AIP roster against online databases maintained by the Women's College Coalition,\footnote{http://www.womenscolleges.org} the White House Initiative on HBCUs,\footnote{https://sites.ed.gov/whhbcu} and the Hispanic Association of Colleges \& Universities.\footnote{https://www.hacu.net}

The database had {a total of} 154 entries {(Table~\ref{tab:demo})}. Each entry in the database included information about the university and department per the Carnegie classification system~\cite{Carnegie2015} and the AIP roster~\cite{AIP2016}, respectively. {We} used publicly available demographic data to determine whether or not a given college or university was a Predominantly White Institution (PWI), i.e., whether or not white students comprised more than 50\% of the student body. Finally, we added contact information for department chairs and relevant lab instructors; this information was publicly available on department websites.

We solicited participation by email. We contacted everyone in the database for whom we had contact information: 150 department chairs, 64 optics instructors, and 62 electronics instructors. In our email solicitations, we specified that we were interested in discussing lab courses with at least one activity that used a photodiode or op-amp circuit. In total, 35 instructors from 27 unique institutions participated in our study: 19 optics instructors from 16 institutions, and 16 electronics instructors from 16 institutions. Physics departments included small (up to 5 physics bachelor's degrees per year), medium (10 to 30 degrees per year), and large (60 to 100 degrees per year) undergraduate physics programs. None of the instructors in our study taught at an HBCU, and none of the optics instructors taught at a {Women's} College. Descriptive information for the corresponding universities is provided in Table~\ref{tab:demo}. 

With respect to the lab courses that were relevant for our investigation, most instructors had multiple years of teaching experience: 18 instructors had taught the course 3 to 10 times and 11 had taught it more than 10 times. The remaining 6 instructors had previously taught the course only 1 or 2 times. At the end of each interview, we asked participants to self-report their race, ethnicity, and gender (question {26} in {the Appendix}). One participant was Black and African American, one was African American and Caucasian, one was Indian, one was Asian, and one was Caucasian with some Asian background. The other 30 participants were white or Caucasian alone, 6 of whom specified European ancestry. Five participants were women, and 30 were men. Four participants identified as cisgender. We do not report intersections of race and gender in order to protect the identities of our research participants.

\subsection{\label{sec:context}Course contexts}
\tabCOURSES
Because our ultimate aim is to develop modeling assessments that are compatible with labs at a range of institutions, it is important to understand the types of courses represented in our study. Accordingly, we describe the content and size of these courses, as well as the background of students who typically enroll in them; {a summary is provided in Table~\ref{tab:courses}}. Information about course context was provided by instructors during the interview, as discussed in Sec.~\ref{sec:collection}.

\subsubsection{Optics labs and related courses}\label{sec:opticslabs}

{Nineteen optics labs were represented in our study: eight optics labs, five advanced labs, and six other types of labs (e.g., intermediate labs, modern labs, or experimental methods courses). Optics lab characteristics are summarized in the left column of Table~\ref{tab:courses}. Although one course was specifically designed for graduate students, undergraduate students commonly enrolled in the course. In almost all cases, students typically completed at least one non-introductory lab course prior to enrolling in the course that was the focus of our interviews. About half of instructors said that students in their courses had previously taken an electronics lab. Other common courses were modern lab, intermediate lab, or junior lab. In cases where upper-division labs are taught infrequently, students had heterogeneous levels of prior lab experience, making it hard to characterize students' overall preparation. However, only one instructor described a course for which students typically had no preparation other than introductory labs.}

Each instructor described a course that covered multiple topics in optics. The most common topics included {spectroscopy, spectrometry, laser beam propagation, interference, interferometry, diffraction, and other laser phenomena or interactions between light and matter}. Less common optics topics included imaging, magneto-optics, and quantum optics. Several instructors also covered topics related to nuclear physics, general instrumentation, or data analysis. {Many} lab courses emphasized communication skills; such courses required students to create written or oral presentations of their work, and some were specifically designated as writing-intensive courses.

{Instructors described 14 distinct optics activities that incorporate photodiodes.} Five types were each described by two instructors. These activities focused on Fraunhofer diffraction, Fresnel reflection, Gaussian beam propagation, Malus's Law, and photodiode characteristics. Nine other types were each described by a single instructor: laser diode spectroscopy, low-light signaling, Michelson interferometry, optical pumping, plasma spectroscopy, ruby crystal fluorescence, single-photon interference, and the Stefan-Boltzmann Law. {All} activities involved investigating phenomena and models that were consistent with the optics content of the courses more {generally}.

While the target phenomena of the activities discussed during interviews varied from instructor to instructor, many activities had several pieces of equipment in common. All instructors described activities that used photodiodes, light sources, and other optical components (e.g., mirrors, lenses, filters, polarizers, and beam splitters). About half of instructors described activities that also used other types of photodetectors, like a photomultiplier tube or microwave diode detector. Most instructors said they used oscilloscopes, multimeters, dc power supplies, and electric circuits. Lock-in amplifiers and stepper motors were each described by some instructors. Some activities involved illuminating a sample, such as a ruby crystal or rubidium vapor.

\subsubsection{Electronics labs}\label{sec:electronicslabs}

{Sixteen electronics labs were represented in our study: six electronics labs, five analog electronics labs, two circuits labs, two instrumentation labs, and one junior lab. Electronics lab characteristics are summarized in the right column of Table~\ref{tab:courses}. In lower-, mixed-, and upper-division courses, physics majors comprised a minority, a majority, or the entirety of enrolled students, respectively.  For half of mixed- and upper-division courses, prerequisites included one non-introductory lab course: modern physics lab or computational physics lab.}

When describing the topics covered in their courses, instructors typically described the types of circuits students build in the course. Every course covered circuits that included solid state components such as diodes, transistors, and op-amps; common circuits of this type included active filters, amplifiers, and rectifiers. Most courses also covered passive filters consisting only of resistors and capacitors or inductors. {Digital} circuits and logic gates were covered in half of the courses in our study. Common test and measurement equipment included oscilloscopes, multimeters, signal generators, and dc power supplies. Microcontrollers, lock-in amplifiers, and data acquisition systems were each used in a few courses. During lectures or lecture-style components of studio labs, instructors taught dc and ac circuit analysis techniques such as nodal, mesh, and phasor analysis. A few instructors said that they used computer-based simulations to aid in circuit analysis. In several cases, lectures also covered topics from electricity, magnetism, or solid state physics.

\subsection{\label{sec:collection}Data collection}
\tabCODES
We conducted 35 semistructured interviews using a protocol that was designed to probe instructors' perspectives on the role of modeling in optics or electronics activities. {Our interview protocol consisted of {26} questions, which are provided in the Appendix.} The first half of each interview focused on departmental and course context, the second half focused on the details of an optics or electronics activity, and one final question asked about participants' race and gender. Occasionally, the interviewer deviated from the protocol to ask a participant for more information about an idea.

Before transitioning from the first half of the interview to the second, the interviewer asked the interviewee to provide a general overview of a relevant activity (question {11} in {the Appendix}). {Often, instructors described multiple relevant activities. In these cases, the interviewer asked the interviewee to focus on the activity they were most familiar with.} Next, the interviewer provided the interviewee with a digital copy of the Modeling Framework (Fig.~\ref{fig:framework}) {and described the subtasks verbally. Additional details are available in the Appendix}.

Finally, the interviewer transitioned to the second half of the interview, which focused on specific details about a particular lab activity. Even though the questions during this part of the interview focused specifically on activities that use photodiodes or op-amp circuits, interviewees also discussed the roles of models and modeling in their courses more generally.

Interviews were conducted via videoconference, but only audio data were recorded. Each interview lasted 40 to 60 minutes, for a total of about 29 hours of audio data. D.R.D.F. and J.T.S. conducted all interviews. Audio data were transcribed by D.R.D.F., L.R., and J.T.S. The transcripts are the data that we analyzed.

\subsection{Data analysis}
As we have done in multiple previous studies~\cite{Zwickl2015a,Dounas-Frazer2017,Dounas-Frazer2016a,Stanley2017}, we used the Modeling Framework to develop an \emph{a priori} coding scheme to analyze interview transcripts. Our coding scheme is presented in Table~\ref{tab:codes}. For every subtask in the framework, we created a corresponding code category: \textbf{make measurements}, \textbf{construct models}, \textbf{make comparisons}, \textbf{propose causes}, and \textbf{enact revisions}. Most of these categories included \emph{a priori} subcategories that were also informed by the language used in the framework. The only exception to this mapping was for the subtasks related to constructing models. Because we were specifically interested in models of photodiodes and op-amp circuits (see, e.g., research question {Q2}), our coding scheme did not categorize models according to whether they described the measurement or physical system. Instead, our scheme included a single code category related to constructing models, with subcategories corresponding to models of photodiode systems, models of op-amp circuits, models of all other aspects of an activity, and interviewee comments on the divide between measurement and physical systems.

{Our} scheme {included} \textbf{learning goals} as a sixth code category. This code {did} not specifically correspond to a particular modeling subtask or the process of modeling more generally. Rather, it was developed to identify which learning outcomes were deemed important by the instructors in our study. Thus, in alignment with research question Q1, we could determine which modeling subtasks, if any, correspond to learning goals in optics and electronics labs, and we could situate them within the context of other learning goals.

To analyze the interview transcripts, we used a dual-pass approach with two coders: coder 1 and coder 2. For the optics interview data set, D.R.D.F. and B.P. played the roles of coders 1 and 2, respectively. For the electronics interview data set, L.R. was coder 1 and D.R.D.F was coder 2. During the first pass, coder 1 read through each transcript and identified excerpts related to each subcategory. Some excerpts received multiple codes. {For example, one instructor said that it was important to them that students learn how to build circuits and analyze data (\textbf{learning goals}) while also indicating that students are encouraged to engage in data analysis when writing reports (\textbf{make comparisons}):}
\Quote{{\emph{You have to look at that data afterwards and decide what you need to do to draw conclusions. That part I think is important, and to me it's second after they [the students] build the circuit. 'Cause it's a lab class. They have to learn how to build something. So that is what I have them do in the lab period, and I try to use the report-writing period for trying to get them to do more analysis.}}}

Next, for each subcategory, coder 2 read through all the coded excerpts to flag excerpts that did not fit the subcategory.  For each data set, about 1,000 codes were assigned by the first coder across all subcategories. For the optics data, the second coder agreed with 91\% of those assignments; agreement was better than 83\% for the electronics data. Coders 1 and 2 reconciled all discrepancies through discussion. In about half of the cases, the coders agreed that the original code assignment was appropriate. In the other half of cases, a different code was assigned. For example, the following excerpt was originally coded as an example of \textbf{enact revisions}:
\Quote{\emph{They set up the experiment and then they all make the same mistakes, so I let them make the mistakes. Then I point out what the issues they have are.}}
Upon discussion, the coders agreed that the interviewee was not describing changes to an apparatus or model. Instead, the interviewee described an instance where {they} identified students' common mistakes as sources of discrepancy between data and predictions. Accordingly, the excerpt was recoded as an example of \textbf{proposing causes}.

After the two coders reconciled all discrepancies, D.R.D.F. performed a second pass of coding to identify emergent subthemes for each \emph{a priori} subcategory. Subthemes were discussed among the research team as a whole. These emergent patterns are discussed in the next section.

\section{\label{sec:results}Results}

We organize our results into three parts: (A) perceived importance of modeling subtasks, (B) models of photodiodes and op-amps, and (C) learning goals not directly related to modeling. Parts A and B correspond directly to research questions Q1 and Q2. Part C represents emergent information that helps us situate the role of modeling in relation to other features of optics and electronics labs. In all three parts, we address research question Q3 by describing similarities and differences between optics and electronics labs.

\subsection{\label{sec:importance}Perceived importance of modeling subtasks}
\figCHARTS
As shown in Fig.~\ref{fig:charts}, every subtask was identified as important by multiple optics and electronics instructors. Every instructor said at least one subtask {was} important, and about two-thirds (63\%) of instructors listed three or more subtasks as important. Making measurements was identified as important by the largest number of instructors, and enacting revisions by the smallest.

\subsubsection{Make measurements}
Most (80\%) of the instructors in our study said that learning how to make measurements is an important outcome of their course. With respect to this subtask, there were no major differences between optics and electronics instructors. Often, this learning goal was coupled to other learning goals:
\QUOTE{We're dedicated to the fundamentals. It's not that important to us what particular technologies or techniques they learn. What we want, is we want students to learn how to predict, make measurements, understand what things could've gone wrong, and be able to approach new problems.}{Wasabi}
Knowing how to make measurements was also coupled to knowing how to use test and measurement equipment:
\QUOTE{What we're hoping for, as the instructors, is to get them [students] the basic idea about how they can set up a simple instrument to make simple measurements. Our goal is that they can go into a lab and be able to utilize an oscilloscope correctly and to be able to collect meaningful data.}{E08}
Similarly, other instructors said that students should learn ``how to actually use devices," ``to use pieces of equipment that are commonly used in research," or ``to make the equipment function." That is, teaching students how to use measurement devices in order to collect data (and hence perform measurements) was an important goal for most instructors. 

\subsubsection{Construct models}
Over half {(60\%)} of instructors said that one of their course goals was for students to learn how to construct models. Optics and electronics instructors put forth different rationales for valuing model construction generally and, more specifically, the distinction between physical and measurement systems.

In optics, constructing models was often framed as important because models are required to make sense of the results of an experiment or predict the outcome of a future measurement. A few optics instructors noted that, while model construction was a major course learning goal, students do not always construct models of their experiments:
\QUOTE{I think they [the subtasks] are all really important. But probably the one that is hardest for students to remember to do is constructing models. \ldots\ So we make a big deal about, ``You really need to calculate this. Phase space is too big for you to just wander around. You need to understand your system. You have to have a model. If you go and measure something, don't get the measurement and say, `What does that mean?' You should know what measurement you're going to get before you make it, and then wonder why you didn't get it." That's something we emphasize a lot.}{Wasabi}
Some optics instructors said that the distinction between models of the physical and measurement systems was important to them. A few of these instructors said that students in their department had few opportunities in the undergraduate physics curriculum to engage with models of measurement equipment. Therefore, they emphasized modeling the measurement system in their courses. One optics instructor critiqued the distinction between measurement and physical systems because there wasn't a ``sharp distinction in physics" between probe and system.

In electronics, two grain sizes of model were commonly discussed: individual components and whole circuits. {Circuit} models and their external representations were seen as useful tools for characterizing and troubleshooting circuits:
\QUOTE{For me, the point of our class is to make a connection between theoretical models and actual things that you can go and build and measure the characteristics of. \ldots\ You shouldn't see a circuit diagram and just recoil in horror. You should be able to look at it and go, ``Okay, I recognize that component, I recognize that component, I can trace through and see, `Okay, here's where it's likely to be breaking down.'"}{E11}
About half of electronics instructors said that it was important for students to understand how the measurement system works. In particular, they said it was important for students to know how real voltmeters deviate from an idealized model that assumes infinite internal resistance and bandwidth. A variety of physical limitations were {described}, {including input impedance and digitization effects}. A few electronics instructors said they do not teach about nonideal models of the measurement system.

\subsubsection{Make comparisons}
Almost three-quarters (70\%) of instructors said that making comparisons is an important learning outcome for their course. Some optics and electronics instructors framed the act of comparing data to models as a defining feature of physics or science more generally. Similarly, a few instructors said that authentic physics experiments sometimes produce ``weird" results whose validity needs to be checked against theory.

When describing the importance of making comparisons, about half of optics instructors specifically noted that it is important to them that students learn how to analyze data. {Almost} all optics instructors said that students engage in one or more types of data analysis during their lab activities. Most described fitting curves to data, and some described simply plotting data to facilitate visual inspection of trends. Normalizing photodiode output signals, subtracting background signals due to ambient light, and computing averages, variances, reduced chi-squared values of various types of data were each described by some instructors. For example,
\QUOTE{Making comparisons is something we do a lot in our undergraduate labs. Saying whether the difference between measurements and theory is significant or not, what's your standard deviation, what's your expected error. Those are things we really push in undergraduate. I think they're really important as well.}{Cinnamon}
Other optics instructors also noted that uncertainty analysis often plays a role in determining whether the agreement between data and models is ``good enough."  Some said that they require their students to perform sophisticated analyses of error and uncertainty. Some instructors said that rigorous error analysis was not a learning goal for their course, or that they expected students to perform primarily qualitative comparisons.

Most optics instructors said that students make predictions during their lab activities. About half said that students predict the shape of an output signal, such as the positions and relative widths of absorption peaks on a spectroscopy experiment, or the shape of a two-dimensional diffraction pattern from a circular aperture. Some instructors said that students predict the value of a model parameter, like the value of Brewster's angle in a Fresnel reflection experiment, or the wavelength of laser light in a Michelson interferometry experiment. In contrast, some instructors said that their students do not make predictions during their lab activities.

In the context of electronics labs, instructors wanted students to learn how to analyze circuits, make predictions about circuit behavior, and check to see if circuits were performing as expected. {About} half of electronics instructors said that students' comparisons are primarily qualitative, involving no statistics. Only one instructor said that students' analysis and comparisons were primarily quantitative. The following response is typical of these major trends:
\QUOTE{I encourage them to do some sort of quantitative comparison. But at the same time, it's not like it's a statistically rigorous comparison. I usually tell them that if you build a circuit that relies very closely on the gain of, any fixed gain in the system, or anything like that---if there's anything you can't adjust by turning a knob or adjusts itself by feedback or something like that---it's not a good circuit. \ldots\ We're not as meticulous as to have them, say, explicitly measure the resistance of the 1k [input resistor], measure the resistance of the 20k [feedback resistor], figure out what that gain should actually be. I tell them, ``Look it's [the gain is] about 20. And if you're building a real experiment that's all the closest you need to be. It's all the closer you want to have to be."}{E07}
Other justifications for engaging students in primarily qualitative comparisons included appeals to the nature of electronics as a discipline. One instructor referred to electronics as ``a 10\% science," {reasoning that, because resistor tolerances are about 10\%}, it is reasonable for the predicted and observed gain of an amplifier circuit to differ by about 10\%. Another called electronics an example of ``yes-no physics" since circuits were often evaluated in a binary way: working or not working.

\subsubsection{Propose causes}
Half (49\%) of instructors identified proposing causes for discrepancies between data and models as an important learning outcome of their course. Both optics and electronics instructors said that students struggle to propose causes on their own because they are unfamiliar with the nonideal behavior of devices. As a result, students often ask their instructors for help identifying problems. One instructor described proposing causes as ``an instructor-meditated conversation."

For some optics instructors, students were expected only to identify and explain discrepancies in their lab reports:
\QUOTE{I mean that's how we actually evaluate the report. It's not so much on how good of an agreement they get. It's more about can they explain the disagreement. }{Rosemary}
In other cases, optics instructors expected students to try to minimize discrepancies by improving their experiment in some way.

Electronics instructors expressed the importance of teaching students how to propose causes on their own:
\QUOTE{We all make mistakes. But as you get better at this and you get more experience, you're going to have to learn to find those problems yourself. Because, if you start working in research lab as a grad student, I'm not going to be there."}{E16}
Electronics instructors said that, when asked for help, they suggested potential causes and solutions, coached students through the process of identifying causes, and asked students to explain the nature of the problem and their attempts to diagnose or fix it. These practices closely resemble those that we characterized in a previous study of electronics instructors' approaches to teaching students how to troubleshoot circuits~\cite{Dounas-Frazer2017}.

\subsubsection{Enact revisions or other iterations}
A third (34\%) of instructors identified enacting revisions or other iterations as important. Both optics and electronics instructors articulated a tension between the time required to iteratively improve an experiment and the limited amount of time available to students in the lab.

About half of optics instructors said that it is important for students to learn about the iterative nature of modeling and experimentation. For example,
\QUOTE{The other thing I really liked about [the Modeling Framework] is the idea of iteration. One of the biggest changes that I see---the positive changes that I see---in the students, is that they go from having a very static, fixed view of everything, that like, ``Oh, this should all be working because I'm taking a class, and it will always work." But then realizing that they need to be constantly checking and revising their understanding of the experiment and the model that they developed for how things work.}{Sage}
Common apparatus revisions included realigning optics, adding or removing optical components, revising the photodiode circuit, blocking ambient light, and changing settings on equipment. Sometimes, apparatus revisions were made in the context of troubleshooting problems. Model revisions included idiosyncratic changes to models of phenomena that were specific to a particular experiment. Such revisions also involved fixing computational mistakes or accounting for nonideal aspects of photodiodes (e.g., finite active area, nonzero response time, or nonlinear voltage responses at very low or high light intensities). In contrast, some optics instructors said that engaging students in iteration was not a goal of their activities. We have published a more thorough analysis of optics instructors' perceptions of revision and iteration elsewhere. Interested readers can find more details in Ref.~\cite{Dounas-Frazer2017arXiv}.

Almost all electronics instructors described examples of students revising apparatus, and about half described examples of students revising models. Students typically revise their circuits due to poor construction. Poor construction practices include wiring the circuit incorrectly, using the wrong components, or using components that don't work. {Other commonly articulated revisions include adjusting connections between the circuit and the oscilloscope and decreasing the amplitude of their input signals due to saturation effects.} When electronics instructors described student revision of models, it was typically in the context of addressing limitations of idealized models of components, circuits, or equipment:
\QUOTE{It's usually the fact that we overly simplify the components in the circuit, so they [the students] don't need usually to redo a measurement. What they need to do though is try to reconcile with this new information about the approximation, they need to reconcile their measurement with the new news about the components. And then the test of that is, when they make another measurement and are seeing something flaky, to try on their own to figure out if this new information about the devices---the less idealized information---will also answer that.}{E04}
A few electronics instructors explicitly said that students do not engage in model revision in their courses.

\subsection{\label{sec:equipment}Models of photodiodes and op-amps}
\tabMODELS
Optics and electronics lab instructors were asked to describe the types of models students use when working with photodiodes and op-amps or inverting amplifiers. Table~\ref{tab:models} provides a summary of the model components identified as important by instructors.

\subsubsection{Models of photodiodes}

Most {optics} instructors said that students modeled the photodiode measurement system as a linear device for which output current or voltage is directly proportional to incident light intensity. Photodiodes were typically used to measure how changes to the experimental setup result in changes in light intensity. For almost all activities, the raw data included output voltages of a photodiode circuit attached to an oscilloscope, multimeter, or lock-in amplifier. Other types of raw data included measures of distance, angle, time, or wavelength, as measured by rulers/micrometers, protractors, clocks, or spectrometers. In almost all activities, the photodiode signal was a dependent variable that changed as a function of other parameters.

About half of {optics} instructors said that students needed to know that photodiodes are linear only within a finite dynamic range, i.e., when the output signal is above a noise threshold and below a saturation threshold. {Another commonly discussed limitation was} the finite size of the photodiode's active area. According to instructors, the active area is an important consideration when focusing light onto the detector or accounting for the photodiode's nonzero internal capacitance and corresponding finite response time.

{About half of optics instructors said that students needed to know about band gaps, \emph{p-n} junctions, and other solid state concepts only ``at a real low level."}  For instance, one instructor said that students need to know only the most basic mechanism of operation: that incident light liberates charge carriers. When the system is not behaving as expected due to, e.g., saturation effects, instructors noted that more sophisticated models are needed. {One} instructor who described an activity focused on characterizing photodiodes said,
\QUOTE{Initially, we treat it [the photodiode] like an ideal source. And then we introduce more complication. Kind of our most complete model, we treat it like an ideal diode which saturates, with a capacitance.}{O12}

{Other features of photodiodes were also relevant in many activities.} In some cases, the spectral range of the photodiode was important. However, most instructors described activities that used a monochromatic light source or a light source whose wavelength changed by only a negligible amount. Hence, in most cases, the spectral range of the photodiode was not important. {A} few instructors said that it was important for students to calibrate their photodiode output in order to compute an absolute light intensity. In almost all activities, only relative intensities were important, and hence output current or voltage was not converted to intensity.

\subsubsection{Models of op-amps and op-amp circuits}
In {electronics} labs, students typically measure the amplitude, frequency, phase, and qualitative aspects of the waveform (e.g., sinusoidal versus triangular) of electric input and output signals. They also measure {resistance and current}.

About half of {electronics} instructors said that they treat op-amps as black boxes, providing students with at most a cursory description of the devices' internal structure. About half said that they describe the internal structure in detail, saying that it was important to ``demystify the black box" and show students why, e.g., op-amps need to be powered by an external power source. One instructor described an activity in which students build an op-amp out of transistors; the others said they covered the internal structure of op-amps during lectures.

Almost all {electronics} instructors said that students use ideal models of amplifier circuits: they are circuits whose output is linearly proportional to the input with a gain determined only by resistor values. Most said that students used one or more of the following properties of closed-loop op-amp circuits: (i) there is no voltage difference between the inverting and noninverting op-amp inputs; (ii) the op-amp inputs have infinite input impedance, and therefore no current flows into the inputs; and (iii) the op-amp output has zero output impedance, and therefore the output voltage remains constant even if the output current changes. Principles (i) and (ii) are often referred to as \emph{the golden rules for op-amps}~\cite{Horowitz1989}. Additionally, about half of instructors said that it was important for students to understand the concept of feedback and its role in closed-loop circuits.

The following excerpt is typical of electronics instructors' discussion of models of op-amps and op-amp circuits:
\QUOTE{The way that I presented the op-amp, in class at least, I first show them a diagram of sort of a simple op-amp in terms of discrete transistors. And I tell them, ``Listen, we could spend the rest of this semester trying to understand what's going on in here, but at some point there's a practicality involved in electronics where you just sort of say, `Well, okay, I know how this thing is supposed to work, right?' And trust that it does." So that's when we end up kind of moving toward, ``An op-amp is something of a black box with your specific rules. Your golden rules of op-amps: the input draws no current and, if you have negative feedback, it works to make both inputs the same." And I sort of tell them, ``Listen, if you have negative feedback, those are pretty much the two rules. And the output can swing to the rails."}{E07}
Here, ``swing to the rails" is a jargon phrase that refers to the phenomenon of saturation.

Other instructors also noted that students encounter physical limitations of amplifier circuits. Saturation, bandwidth, and slew rate issues were each described by about half of the instructors in our study. {Some} instructors said that students encounter small nonzero voltage differences across op-amp inputs, current flowing into the op-amp inputs, or output impedance at the op-amp output. These phenomena are not explained by the ideal model of an amplifier circuit. Rather, they inform the parameter regime in which the ideal model is applicable (e.g., the input voltage has to be sufficiently small that the amplified output does not cause the circuit to saturate).

{For} external representations of models, most instructors said that students use diagrams, schematics, or datasheets. Some said that students create Bode plots. In these cases, Bode plots were used to empirically identify the cutoff frequency of a circuit, i.e., the frequency of input signals above which bandwidth limitations of the op-amp cause it to deviate from the ideal linear model.

\subsection{\label{sec:learninggoals}Learning goals not directly related to modeling}
{Although our study was designed to elicit information about the role of models and modeling in optics and electronics labs, our analysis of emergent themes uncovered other common learning goals of these courses}.

\subsubsection{Written communication skills}

About half of optics and electronics instructors said that developing students' written communication skills is a major focus of their course. Writing assignments included lab reports and lab notebooks. For example,
\QUOTE{I want them to learn how to document well the work that they're doing. So, good logbook hygiene, if you wanna use those words. \ldots\ I think the thing that prevents students---and this is why I place such a big emphasis, when I teach the course---the thing that prevents students from doing the iterative process is that they're very bad about keeping notes in the logbook.}{Sage}
This instructor perceived value in the formative aspects of notebooks and explicitly connected students' ability to keep good notes to their ability to iteratively improve their experiments. This suggests that developing students' communication and modeling skills need not be thought of as separate learning goals.

\subsubsection{Optics labs: Experimental design skills}

About half of optics instructors said that developing students' experimental design skills is an important goal of their course:
\QUOTE{What I try to put emphasis on is actually that the students actually set up things, that experimental apparatus are not given necessarily to them. So they have some design phase, or at least some `align phase' if you're talking optics. ``Here, this is what we wanna do. Okay. Here. This is the mirrors and the laser. And now, go an' do." That would probably also not necessarily fit too well into this framework.}{Rosemary}
In particular, this optics instructor noted that experimental design is not represented in the Modeling Framework, an observation that is consistent with the limitations of the framework.

{The} following learning goals were each articulated by some optics instructors: familiarity with experimental methods (e.g., software interfacing or light manipulation), positive attitudes about experimentation, and understanding of optics content. Other goals were each identified by a few instructors: time management skills, competence with troubleshooting, ability to work independently in the lab, and comfort with ``fiddling around and seeing weird stuff."

\subsubsection{Electronics labs: Building circuits that work}
One major subtheme that emerged from this study was the idea that electronics labs teach students how to \emph{build circuits that work}. Building circuits that work encompasses design, construction, and troubleshooting skills. Almost all electronics instructors said that one or more of these aspects of building {functional} circuits was an important learning goal for their course:
\QUOTE{They learn some basic ideas about designing simple circuits that will do simple things. \ldots\ They have to show that it works. They have to make it work. \ldots\ One thing that I think is very important is learning how to, like, put something together to make it work.}{E01}

Specifically, most electronics instructors said that developing students' troubleshooting skills was {important. Some} noted connections between troubleshooting and the modeling process:
\QUOTE{Like, predict, compare, propose causes, and enact revisions, right? That's the troubleshooting process, which is one of the main goals.}{E12}
This electronics instructor framed troubleshooting as the context in which students enact revisions to their circuits, and making the circuit work was framed as the purpose. Multiple subtasks were identified as comprising the troubleshooting process, a main learning goal for the instructor's course. These findings suggest that supporting students to build circuits that work and developing their modeling ability are compatible learning goals.

{Affective} learning outcomes (e.g., confidence, independence, or perseverance) were identified as important goals by about half of electronics instructors.

\section{\label{sec:discussion}Discussion}
The purpose of our investigation was to align the design of lab-based modeling assessments with the self-reported learning goals of optics and electronics lab instructors. In this section, we discuss our study's (A) limitations, (B) research questions, (C) implications for assessment development, and (D) implications for {future} research.

\subsection{Limitations}
Three limitations must be taken into account when interpreting the results presented here. First, all of the people who participated in our study did so voluntarily and with no monetary incentive. Moreover, we did not perform any classroom observations, nor did we collect any instructional artifacts. Therefore, our findings are likely biased toward the experiences of instructors who enjoy reflecting upon and discussing their teaching, and there may be mismatches between participants' articulated versus actual {teaching practices}.

Next, although the course contexts {appear} to be typical of optics and electronics labs---an important consideration when generalizing from qualitative research~\cite{Eisenhart2009}---almost all participants were men, most were white men, {most taught at PWIs, and none taught at an HBCU}. Accordingly, the personal and institutional priorities and constraints of some populations of lab instructors are not represented in our results.

Last, in contrast to one of the trends in K-12 science education research~\cite{Krell2016,Fazio2012,Justi2003,VanDriel1999,Crawford2004,Davis2010}, our study was not designed to probe instructors' metamodeling knowledge. Rather, it was designed to {determine} whether and how instructors value or implement various modeling subtasks in their classrooms. As a result, we cannot make statements about what instructors think modeling \emph{is}. Nevertheless, we {can make some claims about} what instructors think modeling \emph{is for}~(cf. Refs~\cite{Giere2009,Gouvea2017}). For example, in electronics labs, models help students build circuits that work.

Despite these limitations, the results of this study provide useful insight into the role of modeling in physics labs, which in turn has implications for the design of modeling assessments in optics and electronics.

\subsection{Research questions}
To inform the creation of test objectives that are relevant to instructors and test item contexts that are familiar to students, we set out to answer three research questions:
\begin{itemize}
\item[Q1.] According to instructors, which subtasks of the Modeling Framework are important for students to learn during (a) optics lab courses and (b) electronics lab courses?
\item[Q2.] According to instructors, how do students model (a) photodiodes in optics labs and (b) op-amps and op-amp circuits in electronics labs?
\item[Q3.] What similarities and differences exist, if any, in the ways that models and modeling manifest in optics labs as compared to electronics labs?
\end{itemize}

{For} Q1 and Q3, we found that all modeling subtasks were perceived to be important by multiple instructors in our study, and about two-thirds of instructors listed three or more subtasks as important. In particular, making measurements, constructing models, and making comparisons were each identified as important by a majority of interviewees. Enacting revisions was identified as important by fewer than half of instructors. Optics and electronics instructors alike indicated that knowing how to make measurements requires familiarity with the operation of measurement equipment, comparing data to predictions is inherent to the practice of physics, students struggle to propose causes for discrepancies, and time constraints make it difficult to engage students in the process of revising an experiment. One major difference between optics and electronics instructors is that the latter often framed models and modeling in the context of building {functional} circuits. In electronics, instructors noted that comparisons often amount to qualitative checks of circuit performance. {In contrast, optics instructors more frequently described rigorous data analysis procedures}. 

Consistent with other work~\cite{Dounas-Frazer2017}, most electronics instructors said that troubleshooting was a major learning goal. However, even though diagnosing and repairing a circuit necessarily involves proposing causes and enacting revisions~\cite{Dounas-Frazer2016a}, these modeling subtasks were identified as important by fewer instructors than all other subtasks. {This suggests that some instructors may view troubleshooting as distinct from modeling, potentially because it is more strongly associated with circuit construction.}

{For} Q2 and Q3, we found that photodiode measurement systems and op-amp amplifier circuits were commonly modeled as black boxes whose output signal is linearly proportional to the input signal. In the black box view, photodiodes convert light intensity into an electric output (either current or voltage) with a scalar conversion factor that is often unknown. For op-amp circuits, on the other hand, the scaling factor is typically determined by the values of resistors in the circuit. In both cases, deviations from the black box model included nonlinear effects that arise due to saturation (input is too large), noise thresholds (input is too small), or bandwidth limitations (input changes too quickly). Our results suggest that solid state physics models of photodiodes and op-amps are rarely used in optics and electronics activities, though about half of electronics instructors {reported discussing} the internal structure of op-amps at the transistor level. Use of diagrams, schematics, and data sheets was described more often in the context of working with op-amps and op-amp circuits than with photodiode measurement systems.

\subsection{Implications for assessment development}
{The work presented here is part of a broader effort to create assessments of students' experimental modeling skills and report on the development process, per recent calls from the National Research Council~\cite{DBER2012,NRC2013}. Specifically, research questions Q1, Q2, and Q3 were developed as part of the first phase of a four-phase assessment development process. Nevertheless, they have implications for all four phases. Here, we describe our work's implications for each phase of assessment development.}

\subsubsection{{Results of Phase 1}}
{The main objectives of the first phase are to identify assessment contexts and test objectives that will likely be relevant to a broad range of instructors. By ``assessment contexts," we mean the physical phenomena and experimental systems that are being modeled.} Other studies have shown that students' construction and evaluation of models depend on their domain knowledge~\cite{Ruppert2017,Stewart2005,Fortus2016,Zwickl2015a}. Therefore, assessments of model-based reasoning should be contextualized by phenomena and systems with which students are familiar. Along these lines, {our study provides evidence to support the following suggestions about} the content of future modeling assessments in optics and electronics labs.

In optics labs, experimental systems consisting of laser light, lenses, mirrors, filters, polarizers, photodiodes, oscilloscopes, and multimeters are likely familiar to a wide range of student populations. Given that photodiode measurement systems produce electric signals, it is likely that basic concepts like current, voltage, and Ohm's Law are familiar to these students. However, only about half of instructors said their students had previously completed electronics courses {(Sec.\ \ref{sec:opticslabs})}. Hence, it is unwise to assume widespread familiarity with more advanced electronics concepts. Therefore, simple setups that involve measuring changes in laser light by monitoring the output voltage of a photodiode measurement system are good candidates for contextualizing an experimental optics modeling assessment.

In electronics labs, analog systems that consist of simple active circuits (e.g., op-amps, resistors, capacitors, and inductors) and use oscilloscopes, multimeters, signal generators, and dc power supplies are likely familiar to many student populations. For many students, the electronics lab may be their first lab course {beyond the introductory sequence} {(Sec.\ \ref{sec:electronicslabs})}. Accordingly, electronics modeling assessments should be contextualized only by equipment, concepts, and procedures that are typical of electronics labs.

As for test objectives, our results suggest that {assessments of experimental modeling skills} should assess students' competence with most of the subtasks of the Modeling Framework. There is no clear evidence for designing an instrument that targets only one particular subtask; neither is there strong evidence for excluding any particular subtask. {This suggestion aligns with findings of previous research on students' approaches to modeling in experimental contexts, which are summarized in Sec. \ref{sec:applications}. Namely, multiple studies have demonstrated that students engage in a variety of modeling subtasks when completing experimental optics and electronics tasks~\cite{Zwickl2014perc,Zwickl2015a,Dounas-Frazer2015,Dounas-Frazer2016a}.}

In order to have broad relevance, instruments should assess students' competence with modeling subtasks that many instructors perceive to be important to learn or difficult to master. Assessments of difficult-to-master subtasks could be especially useful for identifying instructional contexts that successfully improve students' proficiency with challenging aspects of modeling. Making measurements and comparisons were both identified as important by a majority of instructors, and proposing causes was often characterizing as hard to learn. Therefore, these subtasks are good candidates for the focus of future modeling assessments.

\subsubsection{{Implications for subsequent phases}}

{The contexts and objectives of an assessment help determine all other details of the instrument. Hence, the results from the first phase of assessment development inform all subsequent phases.}

{In the next phase, we will characterize student navigation of lab practicum-style activities using think-aloud problem solving methods. We have designed optics and electronics lab practicum-style activities that will provide additional insight into students' approaches for making comparisons and proposing causes. These activities use equipment and experimental contexts that are common in optics and electronics labs (Sec.~\ref{sec:context}). To this end, we have already reported preliminary results from the electronics activity; R\'ios et al.~\cite{Rios2018arXiv} found that some students continuously make measurements and enact revisions when they cannot propose a cause for an observed discrepancy in circuit performance. In ongoing work, we are continuing to collect and analyze student data in experimental optics and electronics contexts as part of the second phase of our assessment development plan.}

{In the third and fourth phases, we will create free-response and multiple-response instruments that can be administered at scale. Because the second phase of the process is still ongoing, the scope of claims we can make about the final phases is more limited. Our results suggest that assessments of students' model-based reasoning in upper-division optics and electronics labs should assess most modeling subtasks, especially making measurements and comparisons; they should be contextualized by simple setups (e.g.,  lasers and lenses in optics; analog circuits and oscilloscopes in electronics) and simple component models (e.g., linear response, saturation, and bandwidth limitations); and test items should be designed to elicit students' justifications for choosing particular modeling pathways.}

\subsection{Implications for {future} research}
The investigation presented here is the second national interview study of instructors' perspectives on learning goals and other aspects of teaching upper-division physics labs; a previous investigation focused on electronics lab instructors' ideas about, and approaches to, teaching students how to troubleshoot electric circuits~\cite{Dounas-Frazer2017}. Given the myriad of learning outcomes for undergraduate physics labs~\cite{AAPT2015}, such investigations are important because they help us understand how particular learning goals are valued in different course contexts. For example, our work allows us to use the Modeling Framework as a lens for understanding the global versus domain-specific nature of modeling in physics labs. This and previous work~\cite{Dounas-Frazer2017} demonstrate that the ability to troubleshoot is a primary goal for electronics labs. However, only a few optics instructors in our study specifically named troubleshooting as a learning goal for their courses. {Some indicated that it is sufficient for students to explain empirical results that are at odds with theoretical expectations, but not necessarily revise the experiment}. This contrast may be due to the ease and speed with which components of electric circuits can be replaced or rearranged compared to the relatively tedious and time consuming process of realigning a revised optical setup. {Further} research is needed to understand whether and how the importance of troubleshooting and modeling varies across subdisciplinary domains.

Additionally, national interview studies can help us understand how different learning goals manifest in a particular course context. For example, experimental design skills are important in optics labs, circuit construction skills are important in electronics labs, and developing written scientific communication skills is a common learning goal in both types of courses. While communication, design, and construction may seem distinct from modeling, modeling nevertheless plays an important role in all of these aspects of science~\cite{Koponen2007,Passmore2012,Gilbert2016a}. Future work could explore whether and how the Modeling Framework maps onto these practices, though it was not originally developed to do so (Sec.~\ref{sec:scope}).  Given that these practices are important to many optics and electronics lab instructors, upper-division physics labs provide a promising environment for such investigations. Indeed, we have already mapped the Modeling Framework onto some communication- and design-oriented learning outcomes in undergraduate labs, namely, maintaining lab notebooks~\cite{Stanley2017} and troubleshooting electric circuits~\cite{Dounas-Frazer2016a} in electronics labs.

\section{\label{sec:summary}Summary and future directions}

This study addresses national calls for increased research in labs, including a focus on the process of developing instruments that assess experimental physics practices~\cite{DBER2012,NRC2013}. {To meet these calls}, we presented the first phase of a four-phase process for designing assessments of students' experimental modeling skills, a nationally recognized learning outcome for undergraduate physics labs~\cite{AAPT2015}. The primary goal of this study was to inform the development of test objectives and contexts that are relevant to optics and electronics lab instructors, thus increasing the likelihood that we will develop assessments that many instructors find useful and valuable in their particular domains. 

{We} interviewed 35 instructors about their perceptions of the role and importance of models and modeling in optics and electronics labs. The Modeling Framework for Experimental Physics informed the design of both the protocol that we used to conduct interviews and the \emph{a priori} coding scheme that we used to analyze interview transcripts. We found that each subtask of the modeling framework---making measurements, constructing models, making comparisons, proposing causes, and enacting revisions---was perceived as an important learning outcome by multiple interviewees. Future work will {advance the next phases} of our assessment development process: successively designing hands-on lab practicum-type activities, free-response assessments, and, ultimately, standardized and scalable {CMR-like} assessments of students' model-based reasoning in experimental optics and electronics contexts.

The creation of scalable modeling assessments for optics and electronics lab courses will facilitate the research and development of instructional labs in two ways. First, it will broaden the landscape of instruments available to researchers and instructors engaging in research-based transformation of upper-division lab courses. Second, scalable modeling assessments with centralized administration will open the door to nationwide studies of lab courses (cf. Ref.~\cite{Wilcox2016b}). {Such studies will} help identify which instructional strategies are effective at improving students' modeling abilities in experimental physics contexts.

\begin{acknowledgments}
We are grateful to the instructors who volunteered to participate in this study. This material is based upon work supported by the NSF under Grant Nos. DUE-1611868, DUE-1726045, and PHY-1734006.
\end{acknowledgments}

\appendix*
\section{{Interview protocol}}
\begin{enumerate}[noitemsep]
\item {How many majors graduate from your department each year?}
\item {How many lab courses are offered at your institution?}
\item {For this interview, let's focus on the course you're most familiar with. How many times have you previously taught this course?} 
\item {Without going into the details of requirements for different major tracks, can you tell me which other lab courses students typically complete before enrolling in the course?}
\item {How many students are typically enrolled in this course?}
\item {Are the students who take this course typically in their first, second, third, or fourth or higher year of college?}
\item {Approximately what fraction of students in this course are physics majors?}
\item {What are the main things students should learn in this course?}
\item {What topics do you typically cover in this course?}
\item {Is there anything special about this course that you want to tell me about?}
\item {Specifically, I'm curious about any activity that uses \textless photodiodes or op-amp circuits\textgreater. Can you briefly describe for me any activity that meets that criterion?} 
\end{enumerate}

{After the interviewee responded to question 11, the interviewer showed them a digital copy of the Modeling Framework (Fig.~\ref{fig:framework}). When introducing the framework, the interviewer read the following script:}
\Quote{{The way my research group has been thinking about activities like this is by using something we call the Modeling Framework. I want to understand how this framework applies to your activity, so the rest of this interview will focus on the framework. Before I move ahead, I want to talk through the framework with you so that you and I are on the same page about the relevant vocabulary and processes. The framework looks complicated, but it's not that bad when we break it down piece by piece.}}
{The interviewer then gave a verbal description of all six subtasks. For example, the interviewer described comparisons as follows:}
\Quote{{The next gray box corresponds to comparing data to predictions. Depending on the results of the comparison, our framework describes three potential outcomes. First, if the measurement is good enough, then the modeling process is done. What it means to have ``good enough" agreement is highly context dependent. If it's difficult to compare the data to a prediction because, say, the data are too noisy, it might be the case that more data need to be collected. This is the ``Maybe" pathway in the diagram. The third and last possibility is that there is an obvious disagreement between the data and prediction, in which case the next two subtasks become relevant.}}

{After describing the subtasks, the interviewer gave the interviewee a chance to ask clarifying questions about the framework. Then, the interviewer moved on to the second part of the interview.}
\begin{enumerate}[resume,noitemsep]
\item {When working on this activity, what equipment do students use?}
\item {What theoretical principles or concepts do students need to know in order to successfully complete the activity?}
\item {What limitations, assumptions, and simplifications are necessary to successfully complete the activity?}
\item {When working on the activity, what equations or diagrams do students use?}
\item {What predictions do students make when working on the activity?}
\item {What types of measurements do students perform when working on the activity?}
\item {What types of data do students collect when working on the activity?}
\item {What types of analyses, calibrations, or unit conversions do students perform on their raw data?}
\item {In what ways do students compare their predictions to their data  when working on the activity? For example, do they make qualitative comparisons, order of magnitude comparisons, or use statistical tests?}
\item {Do students have opportunities to collect additional data if they aren't sure whether their data and predictions agree? If so, can you tell me more about this?}
\item {Do students propose explanations for why their data and predictions don't agree? If so, can you tell me more about this?}
\item {Do students enact revisions to the equipment, apparatus, or models? If so, can you describe them for me?}
\item {Now that I have a better understanding of what students do, I'd like to get a better picture of what you think is important for students to learn about experimental physics. Which components of the Modeling Framework are the most important for students to learn and which are the least important?}
\item {Is there anything else you'd like to tell me?}
\item {I'm trying to reach out to a broad range of instructors from different institution types to be sure I collect diverse perspectives about modeling. For the transcript record, is it okay if I ask about your gender, race, and ethnicity?}
\end{enumerate}


\bibliography{./modeling_test_objectives_database}

\end{document}